\documentstyle[aps,pra,multicol,graphicx,eqsecnum]{revtex}
\begin{document}
\draft
\title{Effects of twin-beam ``squashed'' light on a three-level atom}
\author{L.K. Thomsen and H.M. Wiseman}
\address{School of Science, Griffith University, Nathan 4111, Australia}
\date{\today}
\maketitle

\begin{abstract}
An electro-optical feedback loop can make in-loop light (squashed
light) which produces a photocurrent with noise below the standard
quantum limit (like squeezed light). We investigate the effect of
squashed light interacting with a three-level atom in the cascade
configuration and compare it to the effects produced by squeezed
light and classical noise. It turns out that one master equation
can be formulated for all three types of light and that this
unified formalism can also be applied to the evolution of a
two-level atom. We show that squashed light does not mimic all
aspects of squeezed light, and in particular it does not produce
the characteristic linear intensity dependence of the population
of the upper-most level of the cascade three-level atom.
Nevertheless, it has non-classical transient effects in the
de-excitation.
\end{abstract}
\pacs{42.50.Dv, 42.50.Ct, 32.80.-t}

\newcommand{\beq}{\begin{equation}}
\newcommand{\eeq}{\end{equation}}
\newcommand{\bqa}{\begin{eqnarray}}
\newcommand{\eqa}{\end{eqnarray}}
\newcommand{\nn}{\nonumber}
\newcommand{\nl}[1]{\nn \\ && {#1}\,}
\newcommand{\dg}{^\dagger}
\newcommand{\bra}[1]{\langle{#1}|}
\newcommand{\ket}[1]{|{#1}\rangle}
\newcommand{\ip}[2]{\langle{#1}|{#2}\rangle}
\newcommand{\sch}{Schr\"odinger }
\newcommand{\schs}{Schr\"odinger's }
\newcommand{\hei}{Heisenberg }
\newcommand{\heis}{Heisenberg's }
\newcommand{\bl}{{\bigl(}}
\newcommand{\br}{{\bigr)}}
\newcommand{\mod}[1]{\left| {#1} \right|}
\newcommand{\sq}[1]{\left[ {#1} \right]}
\newcommand{\cu}[1]{\left\{ {#1} \right\}}
\newcommand{\ro}[1]{\left( {#1} \right)}
\newcommand{\an}[1]{\left\langle {#1} \right\rangle}
\newcommand{\implies}{\Longrightarrow}

\begin{multicols}{2}

\section{Introduction}

Squeezing is a form of non-classical light which involves the
reduction of noise in one quadrature below the standard quantum
limit, and the increase of the noise in the conjugate quadrature
\cite{WalMil94}. However the characteristic below-shot-noise
photocurrent is not limited to squeezed light. An electro-optical
feedback loop can also produce an (in-loop) homodyne photocurrent
with a below-shot-noise spectrum. The resultant in-loop light is
called ``squashed'' light \cite{Bucetal99,Wis99,why}. Our interest
lies in understanding the non-classical effects of the interaction
of squashed light with atoms.

It has recently been shown \cite{Wis98} that illuminating a
two-level atom with squashed light causes line narrowing of one
quadrature of the atom's fluorescence, an effect also produced by
squeezing \cite{Gar86}. This effect of squeezing (or squashing)
has not yet been observed \cite{Turetal98}. However, it has been
demonstrated \cite{3laexpt95} that a squeezed two-photon field
exciting a three-level atom in the cascade configuration gives an
excited population for the highest level with a linear as well as
quadratic (classical) dependence on intensity. This effect was
first predicted in Ref.~\cite{FicDru91} and the experiment has
recently been analyzed by the same authors \cite{FicDru97}. This
experiment motivates our current work to see if squashing also
produces effects similar to squeezing when illuminating a cascade
three-level atom.

Although based on the same techniques as in \cite{Wis98}, the
development of the model for the three-level system required a
complete redesign of the feedback mechanism. Non-classical
excitation of a three-level atom requires non-classical
correlations between two fields, i.e. a twin-beam, which couple to
the two transitions in the atom. Twin-beam correlations can be
produced in a double feedback system, where the fed-back
photocurrents result from linear combinations of the twin-beam
quadratures. These are analogous to the twin-beam squeezing
correlations produced by a non-degenerate optical parametric
oscillator \cite{3laexpt95}. The resultant master equation for a
cascade three-level atom interacting with twin-beam squashed light
can be compared to those for squeezed \cite{FicDru91} and
classical twin-beams.

The structure of the paper is as follows. Section II introduces
the interaction of non-classical light with simple atomic systems.
Specifically, it covers previous results of the squeezed
excitation of a two-level atom, and it summarizes the production
of squashed light and its interaction with a two-level atom. The
techniques and concepts introduced here will be used again in
later sections.

Section III presents the theory for a three-level atom coupled to
a broadband squeezed field, and introduces our definition of a
``twin-beam''. Section IV develops the main theoretical results of
this paper, which include the master equation for a three-level
atom coupled to twin-beam squashed light, and the correlations of
the squashed twin-beam itself.

Section V compares the non-classical nature of squashed light with
squeezed light as well as classical noise. The first two parts of
this section show that a general formalism can be found for all
three types of light interacting with both two-level and
three-level atoms. Finally, Section V also details the specific
classical and non-classical effects of squashed light. Section VI
concludes.

\section{Two-level atom}

The interaction of non-classical light with atoms was first
studied soon after the first incontestible observation of
squeezing \cite{Slu85}. For a recent review of atoms in squeezed
light fields see Ref. \cite{DalFicSwa99}. It began with the
prediction by Gardiner \cite{Gar86} that broadband squeezed light
would produce an arbitrarily narrow line in the power spectrum of
the fluorescence of a two-level atom. This was thought of as a
``direct effect of squeezing'' \cite{Gar86}.

A recent letter \cite{Wis98} by one of us showed that squashed
light also produces a similar line-narrowing, and so this ``direct
effect'' is one of non-classical correlations rather than solely
of squeezing. The work on squashed light spectroscopy of two-level
atoms was recently extended to include simultaneous squeezing and
squashing \cite{Wis99}. Here we review the two effects separately.

\subsection{Squeezed light}

The squeezed vacuum fields that are used as spectroscopic sources are
actually multi-mode squeezed states. The theory of single-mode squeezing
was generalized to multi-mode squeezing by Caves and Schumaker \cite{CavSch85}.
The continuum field is the limit of an infinite number of modes. For
polarized light propagating in one direction, the modes are simply indexed
by the mode frequency $\omega=ck$. The continuum annihilation operators
$b(\omega)$ may be taken to satisfy
\beq
\sq{b(\omega),b\dg (\omega')}=\delta(\omega-\omega').
\eeq
If the occupied modes are restricted to a relatively narrow bandwidth $B$ around
some carrier frequency then, in the time domain, the commutation relation becomes
\beq
\sq{b(t),b\dg(t')}=\delta(t-t'),
\label{commrel}
\eeq
where the width of the $\delta$-function is actually of the order
$B^{-1}$ and $b(t)$ can now be thought of as the annihilation operator
of a ``localized photon''. Thus $b\dg (t)b(t)$ can be interpreted as the
photon flux operator. The $X$ and $Y$ quadratures of the continuum field
are defined as
\beq
X(t)=b(t)+b\dg (t),~~Y(t)=ib\dg (t)-ib(t),
\label{XYdef}
\eeq
and for free fields these obey the commutation relations
\beq
\sq{X(t),Y(t')}=2i\delta(t-t'). \label{commrel2}
\eeq

For fields with zero mean, as are all the fields in this paper, the
$X$ quadrature spectrum is defined as
\bqa
S_{X}(\omega)&=& \langle \tilde{X}(\omega)X(0)\rangle_{\rm ss}, \nn \\
&=&\frac{1}{2\pi}\int_{-\infty}^{\infty}\an{\tilde{X}(\omega)\tilde{X}(-\omega ')}_{\rm ss}d\omega ',
\label{specdef}
\eqa
and similarly for the $Y$ quadrature spectrum. It can then be shown that
for a stationary free field
\beq
S_{X}(\omega)S_{Y}(\omega)\geq 1, \label{hei}
\eeq
which is known as the Heisenberg inequality \cite{Sha87,blurb}.
A vacuum continuum field \cite{Gla63,Sud63} is one such
that for all $\omega$ and all $\theta$,
$S_{Q_{\theta}}(\omega)=1$. Here $Q_{\theta}=
X\cos\theta+Y\sin\theta$. This is called the standard quantum limit or
shot-noise limit. For a \textit{squeezed} continuum field we have
$S_{Q_{\theta}}(\omega)<1$, for some $\omega$ and some $\theta$.

Now consider placing a two-level atom in a squeezed continuum
field. Under the rotating-wave approximation the interaction
Hamiltonian is
\beq
H(t)=i\hbar[b\dg (t)\sigma(t)-\sigma\dg(t)b(t)], \label{H2LA}
\eeq
where $\sigma$ is the atomic lowering operator. Here the degree of
mode matching of the input light modes to the atom's dipole radiation
mode is assumed to be unity and the atomic linewidth is also set to
unity. See Ref.~\cite{Wis98} for details on mode-matching.

Assuming the squeezed light is broadband compared to the atomic
linewidth, we can make the white noise approximation that the
quadrature spectra are constant. For experimental generation of
squeezed light (e.g. an optical parametric oscillator) photons are
produced in correlated pairs with frequencies symmetrically placed
around a center frequency $\omega_{0}$ \cite{DalFicSwa99}. The
photon number function, $N(\omega_{k})=\langle
b\dg(\omega_{k})b(\omega_{k})\rangle$ and the two-photon
correlation function, $M(\omega_{k})=\langle
b(\omega_{k})b(2\omega_{0}-\omega_{k})\rangle$ are both constants
in the broadband approximation.

In general for squeezed light $N\leq |M|\leq \sqrt{N(N+1)}$, where
the upper limit represents the minimum-uncertainty state, also known as
the maximally squeezed state. For minimum-uncertainty squeezing with $M$
real, the light can be characterized by a single real number $L=1+2N+2M$
\cite{WalMil94}. Thus the quadrature spectra are
\beq
S_{X}(\omega)=L=1/S_{Y}(\omega).
\label{L}
\eeq
If $M<0$ then $L<1$ and the $X$ quadrature is squeezed, i.e.
$S_{X}(\omega)<1$, and correspondingly the $Y$ quadrature is stretched,
i.e. $S_{Y}(\omega)>1$.

From these correlations and the interaction Hamiltonian (\ref{H2LA}) one
can obtain the following master equation for the atom \cite{Gar91}:
\beq
\dot{\rho} = \frac{1}{4L}{\cal D}\sq{(L+1)\sigma-(L-1)\sigma\dg}\rho,
\label{2LAsqueeze}
\eeq
where as usual for arbitrary operators $A$ and $B$,
\beq
{\cal D}[A]B \equiv ABA\dg -\frac{1}{2}A\dg AB -\frac{1}{2}BA\dg A.
\eeq

From the master equation the following dynamical equations can be
derived
\bqa
{\rm Tr}[\dot{\rho}\sigma_{x}]&=&-\gamma_{x}{\rm Tr}[\rho\sigma_{x}], \label{blochx} \\
{\rm Tr}[\dot{\rho}\sigma_{y}]&=&-\gamma_{y}{\rm Tr}[\rho\sigma_{y}], \label{blochy} \\
{\rm Tr}[\dot{\rho}\sigma_{z}]&=&-\gamma_{z}{\rm Tr}[\rho\sigma_{z}]-C \label{blochz},
\eqa
where
\beq
\gamma_{x}=\frac{L}{2},~\gamma_{y}=\frac{1}{2L},~\gamma_{z}=\gamma_{x}+\gamma_{y},~C=1.
\eeq
Here we see that, for $X$ quadrature squeezing $(L<0)$, the decay
of $\sigma_{x}$ is below its natural value of 1/2. In terms
of the squeezed spectrum it is given by $S_{X}(\omega)/2$.
This slower decay causes a corresponding line narrowing in the power
spectrum of the atom's fluorescence.

\subsection{Squashed light}

As mentioned in the introduction, there is another approach to producing
light with a sub-shot-noise photocurrent apart from squeezing: this is
squashing. This involves modulating the light incident on a photodetector
by using the very current from that detector, i.e. electro-optical feedback.
Squashed light was first observed \cite{MacYam86} not long after the
first observation of squeezing \cite{Slu85}.

\begin{figure}
\hspace{0.1cm}
\includegraphics[width=0.41\textwidth]{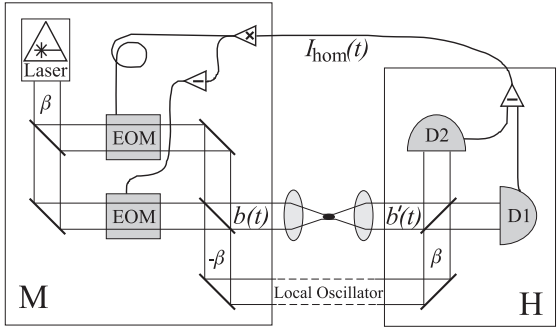}
\vspace{0.25cm} \caption{\narrowtext Experimental configuration
for squashed light incident on a two-level atom. The atom is
represented by a small ellipse at the focus of $b(t)$. The
detection device (region H) finds the difference $I_{\rm hom}(t)$
between the currents at the photodetectors D1 and D2. The
modulation device (region M) amplifies and splits $I_{\rm hom}(t)$
so that two signals (with opposite sign) are fed back to the two
electro-optical modulators (EOM).}
\end{figure}

Consider the experimental apparatus shown in Fig.~1. We begin with
the modulation region, M, which adds a modulated
coherent amplitude to the input vacuum field. There are various
ways of achieving this including the one pictured, which is
detailed in Ref.~\cite{Wis98}. The resultant field, $b(t)$, is incident
on a two-level atom. After the atom, a homodyne measurement is
performed on the field, now given by $b'(t)$. This is done in the
detection region, H, and makes use of the local oscillator also
produced in M. The homodyne photocurrent is then used to control
the modulation of the coherent amplitude, and it is this feedback
that ensures $b(t)$ has noise below the standard quantum limit.

The field exiting the modulator is given by
\beq
b(t)= \nu(t) + \varepsilon(t), \label{b}
\eeq
where $\nu(t)$ represents the input vacuum and $\varepsilon(t)$ is a
real function of time. For a vacuum field all first and second order
moments vanish except for\cite{Gar91}
\beq
\an{\nu(t)\nu\dg(t')}=\sq{\nu(t),\nu\dg(t')}=\delta (t-t'),
\label{vacfluc}
\eeq
and the $X$ and $Y$ quadratures have spectra, defined by (\ref{specdef}),
equal to unity. Thus they are the special case of (\ref{L}) where $L=1$.

For now ignore the two-level atom. The detection device makes a homodyne
measurement of the $X$ quadrature of $b'(t)$, which in the absence of the
atom is equal to $b(t)$. The homodyne photocurrent is then given by \cite{Gar91}
\beq
I_{\rm hom}(t)=X_{\nu}(t)+2\varepsilon(t),
\label{Ihom}
\eeq
where $X_{\nu}(t)=\nu(t)+\nu\dg (t)$ and we have assumed perfect detection
efficiency. Through the feedback loop, this photocurrent is amplified and
used to control the classical field $\varepsilon(t)$. We set
\beq
\varepsilon(t)=\frac{g}{2}\int^{\infty}_{0} h(s)I_{\rm hom}(t-\tau-s)~ds, \label{epsilon}
\eeq
where $\tau$ is the minimum time delay of the feedback, $h(s)$ is the
response function normalized such that $\int^{\infty}_{0}h(s)ds=1$ and
$g$ is the round-loop gain (for stability we require
$g{\rm Re}[\tilde{h}(\omega)e^{-i\omega\tau}]<1$ for all $\omega$) \cite{SteSubWil90}.

Solving equation (\ref{epsilon}) in the Fourier domain, where $I_{\rm hom}(t)$
is given by (\ref{Ihom}), we obtain
\beq
\tilde{\varepsilon}(\omega)
=\frac{g\tilde{h}(\omega)e^{-i\omega\tau}\tilde{X}_{\nu}(\omega)}{2(1-g\tilde{h}(\omega)e^{-i\omega\tau})}.
\eeq
Thus $b(t)$ has an $X$ quadrature spectrum given by
\beq
S_{X}(\omega)= |1-g\tilde{h}(\omega)e^{-i\omega\tau}|^{-2}.
\eeq
At frequencies $\bar{\omega}$ well inside the bandwidth of $\tilde{h}(\omega)$
and much less than $\tau^{-1}$, we are in the limit of broadband feedback
and can therefore set $\tilde{h}(\bar{\omega})e^{-i\bar{\omega}\tau}=1$.
In this case
\beq
S_{X}(\bar{\omega})=(1-g)^{-2}.
\label{spec2LA}
\eeq
Clearly this spectrum is below the standard quantum limit for
negative feedback ($g<0$). For positive feedback, $0<g<1$, we
actually increase the noise.

Also, since the feedback acts only on the $X$ quadrature,
$S_{Y}(\omega)$ remains at the vacuum limit of unity and
\beq
S_{X}(\omega)S_{Y}(\omega)= |1-g\tilde{h}(\omega)e^{-i\omega\tau}|^{-2},
\label{viol}
\eeq
which can clearly be less than one. This apparent violation
of the Heisenberg inequality (\ref{hei}) is due to the fact
that the in-loop light is not a free field. For time differences
$|t-t'|>\tau$, the feedback loop delay time, the usual two-time
commutation relations Eq.~(\ref{commrel}) and Eq.~(\ref{commrel2})
are are no longer valid, which was first shown by Shapiro
{\em et al} \cite{Sha87}. Basically this is because parts of the
field separated by a time greater than $\tau$ don't actually
exist together.

From this we note that non-classical light produced by a
feedback loop is not squeezed light in the ordinary sense
\cite{Wis99}. First, it is not a free field (as explained above),
and second, attempts to remove this light by a beam splitter
actually gives above-shot-noise light, as verified experimentally
\cite{YamImoMac86,WalJak95}. Due to this difference between
in-loop light and ordinary squeezed light, it is termed
``squashed'' light \cite{Bucetal99,Wis99}. As we will show, while
not the same as squeezed light, squashed light does have a
non-classical effect on a two-level atom placed within the
feedback loop.

Return to Fig.~1 and now include the two-level atom. Again assume
that the degree of mode matching of the input light modes to the
atom's dipole radiation mode is unity. Then the interaction
Hamiltonian has the same form as that given in (\ref{H2LA}), and
this linear coupling gives $b'(t)=b(t)+\sigma(t)$ \cite{Gar86}.
Incorporating the feedback as before we find
\beq
\tilde{\varepsilon}(\bar{\omega})=
\frac{g[\tilde{X}_{\nu}(\bar{\omega})+\tilde{\sigma}_{x}(\bar{\omega})]}{2(1-g)}.
\eeq
Again we have assumed the limit of broadband feedback and set
$\tilde{h}(\bar{\omega})e^{-i\bar{\omega}\tau}=1$.

Changing back to the time domain we employ the Markov approximation to
state that the feedback is instantaneous in terms of the characteristic
response rates of the system. This requires $\bar{\omega}$, still well
inside the feedback bandwidth, to be much greater than the atomic linewidth,
i.e. $\bar{\omega}\gg1$. The field incident on the atom is then
\beq
b(t)=\nu(t)+\frac{g[X_{\nu}(t^{-})+\sigma_{x}(t^{-})]}{2(1-g)},
\eeq
where the time argument $t^{-}$ is used to indicate that even under the Markov
approximation the fed-back signal must act after the interaction with the atom.

Substituting $b(t)$ into the interaction Hamiltonian (\ref{H2LA}) gives the Hamiltonian
due to the coupling of the atom to the vacuum, plus a feedback Hamiltonian.
The vacuum Hamiltonian is just that given in (\ref{H2LA}) except $b(t)$ is
replaced by the vacuum operator $\nu(t)$. The feedback Hamiltonian is
\beq
H_{\rm fb}(t)=\frac{\lambda}{2}\sigma_{y}(t)[\sigma_{x}(t^{-})+X_{\nu}(t^{-})],
\eeq
where the feedback parameter $\lambda$ is given by
\beq
\lambda=\frac{g}{1-g}\in(-1,\infty). \label{lambda}
\eeq

To describe the evolution generated by this Hamiltonian, we require the
Heisenberg-picture theory of homodyne detection and feedback \cite{Wis94}.
The result is the following master equation for the atom
\beq
\dot{\rho} = {\cal D}[\sigma]\rho
-i\frac{\lambda}{2} \sq{ \sigma_{y}, \sigma\rho+\rho\sigma\dg }
+ \frac{\lambda^{2}}{4}{\cal D}\sq{ \sigma_{y}} \rho.
\label{2LAsquash}
\eeq
Expressed in terms of $\lambda$, the $X$ quadrature spectrum (\ref{spec2LA})
is given by
\beq
S_{X}(\bar{\omega})=1+2\lambda+\lambda^{2}.
\label{speclamb}
\eeq
Thus we can see that if
$\lambda<0$ we will squash the $X$ quadrature, i.e.
$S_{X}(\bar{\omega})<1$, and the optimal squashing is for $\lambda=-1$.

The master equation again leads to Eqs.~(\ref{blochx})-(\ref{blochz}),
but this time
\bqa
\gamma_{x}&=&\frac{1}{2}(1+2\lambda+\lambda^{2})=\frac{1}{2} S^{X}(\bar{\omega}), \nn \\
\gamma_{y}&=&\frac{1}{2},~~~\gamma_{z}=\gamma_{x}+\gamma_{y},~~~C=1+\lambda.
\eqa
Again the decay of $\sigma_{x}$ is inhibited for negative feedback,
i.e. for $S_{X}(\bar{\omega})<1$. Most importantly, the dependence of
$\gamma_{x}$ on the degree of $X$ quadrature squashing is exactly the
same as for squeezing. The only differences between squeezing and
squashing are that $\gamma_{y}$ is unaffected by squashing and $C$ is
unaffected by squeezing. This follows directly from the squashing of
fluctuations which violates the usual uncertainty relations, as
discussed following Eq.~(\ref{viol}) above.
Comparing the fluorescence spectra for both
cases \cite{Wis98}, it can be seen that they are certainly not identical,
but both have a sub-natural linewidth.

\section{Twin-beam squeezing}

The master equation for a three-level atom in a broadband squeezed
vacuum has previously been developed by Ficek and Drummond
\cite{FicDru91}. We follow a simplified derivation, based on the
same assumptions, which leads to a master equation equivalent to
results derived in Part II, Section III of the above reference.

\subsection{Three-level atom master equation}

Ficek and Drummond \cite{FicDru91} describe the input field as a
broadband squeezed vacuum which is in two-photon resonance with
the atom, i.e. $\omega_{0}=(\omega_{1}+\omega_{2})/2$ (where
$\omega_{1}$ and $\omega_{2}$ are the atomic transition
frequencies). More specifically, they assume the typical squeezing
spectrum of a non-degenerate parametric oscillator where the
signal and idler fields are tuned to resonance with $\omega_{1}$
and $\omega_{2}$ \cite{3laexpt95}. Typically, the squeezing
bandwidth $B$ is small compared with the carrier frequency
$\omega_{0}$.

To simplify the derivation further, we note that if $\omega_{1}$
and $\omega_{2}$ are significantly different, i.e
$|\omega_{1}-\omega_{2}|\gg B$, then the signal and idler fields
can be treated as two (distinct) correlated multi-mode fields.
Hence, we define the input as a broadband, squeezed
\textit{twin-beam} with the two fields defined by respective
continuum annihilation operators $b_{1}(\omega)$ and
$b_{2}(\omega)$. Note the squeezed field is broadband in the sense
that it appears as $\delta$-correlated squeezed white noise to the
atom, i.e. $B\gg\gamma_{1},\gamma_{2}$, which are the atomic
linewidths.

For a broadband squeezed twin-beam with no mean field
$\an{b_{1}(t)}=\an{b_{2}(t)}=0$, while the correlations
are
\bqa
\langle b\dg_{i}(t)b_{j}(t')\rangle &=& N_{i}\delta_{ij}\delta(t-t'),\nn \\
\langle b_{i}(t)b\dg_{j}(t')\rangle &=&(N_{i}+1)\delta_{ij}\delta(t-t'),\nn \\
\langle b_{i}(t)b_{j}(t')\rangle &=&M(1-\delta_{ij})\delta(t-t'),\nn \\
\langle b\dg_{i}(t)b\dg_{j}(t')\rangle &=& M^{*}(1-\delta_{ij})\delta(t-t').
\label{corr}
\eqa
Here $|M|^{2}\leq N_{1}N_{2}+{\rm min}(N_{1},N_{2})= N(N+1)~{\rm
if}~N_{1}=N_{2}=N$, where as before $N$ and $M$ are the broadband
approximations of the photon number function (here a doubly peaked
function) and the two-photon correlation function. Thus $N_{1}$
and $N_{2}$ have the values of the two peaks in the the photon
number function, and correspond to the intensity of each beam.
Note again that the $\delta$-functions in these correlations
are not exact, their widths being actually of the order
$B^{-1}$ (see Sec. II.A).

Consider coupling a three-level atom to the broadband squeezed
twin-beam, where the fields $b_{1}$ and $b_{2}$ are coupled to the
atomic transitions $\ket{1}\rightarrow\ket{2}$ and
$\ket{2}\rightarrow\ket{3}$ respectively. This is shown in Figure
2. Assume again for simplicity that it is possible to mode-match
all of the squeezed twin-beam into the atom's input. The
Hamiltonian of the atom in the interaction picture at time $t$ is
then \bqa
H(t)&=&i\hbar\sqrt{\gamma_{1}}\sq{b\dg_{1}(t)s_{1}(t)-s\dg_{1}(t)b_{1}(t)}
\nl{+}
i\hbar\sqrt{\gamma_{2}}\sq{b\dg_{2}(t)s_{2}(t)-s\dg_{2}(t)b_{2}(t)},
\label{intH} \eqa where the atomic parameters are defined in Fig.~2.

\begin{figure}
\hspace{0.9cm}
\includegraphics[width=0.3\textwidth]{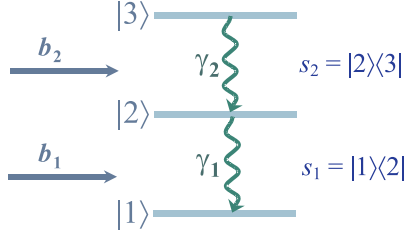}
\vspace{0.25cm} \caption{\narrowtext Coupling of the broadband
twin-beam to the three-level atom. The twin-beam is represented by
the fields $b_{1}$ and $b_{2}$, while $s_{1}=\ket{1}\bra{2}$ and
$s_{2}=\ket{2}\bra{3}$ are the atomic lowering operators and
$\gamma_{1}$ and $\gamma_{2}$ are the spontaneous decay rates from
levels $\ket{2}$ and $\ket{3}$ respectively.}
\end{figure}

Applying the unitary evolution of the Hamiltonian (\ref{intH}) to
an arbitrary system operator $r(t)$ leads to the quantum Langevin
equation for the atom (in the \hei picture). Changing from the
\hei to the \sch picture is achieved using the relation
\beq
\an{dr(t)}={\rm Tr}\sq{d\rho(t)r},
\label{HP-SP}
\eeq
where the picture (\sch or Heisenberg) is specified by the placement
of the time argument. The resulting master equation for the atom is
\bqa
\dot{\rho} &=& (1+N_{1})\gamma_{1}{\cal D}[s_{1}]\rho + (1+N_{2})\gamma_{2}{\cal D}[s_{2}]\rho
\nl{+} N_{1}\gamma_{1}{\cal D}[s\dg_{1}]\rho + N_{2}\gamma_{2}{\cal D}[s\dg_{2}]\rho
\nl{+}
\frac{1}{2}M^{*}\sqrt{\gamma_{1}\gamma_{2}}
 \cu{\sq{s_{1},\sq{s_{2},\rho}}+\sq{s_{2},\sq{s_{1},\rho}}}
\nl{+} \frac{1}{2}M\sqrt{\gamma_{1}\gamma_{2}}
 \{[s\dg_{1},[s\dg_{2},\rho]]+[s\dg_{2},[s\dg_{1},\rho]]\}.
\eqa

If we set $\gamma_{1}=\gamma_{2}=1$, $N_{1}=N_{2}=N$ and $M$ real,
the master equation reduces to
\bqa
\dot{\rho} &=& (1+N)\cu{ {\cal D}[s_{1}]\rho + {\cal D}[s_{2}]\rho }
+ N\{ {\cal D}[s_{1}\dg]\rho + {\cal D}[s_{2}\dg]\rho \}
\nl{+} \frac{M}{2} \left\{ \sq{s_{1},\sq{s_{2},\rho}}+\sq{s_{2},\sq{s_{1},\rho}}
+ [s\dg_{1},[s\dg_{2},\rho]] \right. \nn \\
&&~~~~~~~~~~ \left. +~[s\dg_{2},[s\dg_{1},\rho]] \right\}.
\label{squeeze}
\eqa
The linear versus quadratic dependence of the excited population
on intensity, as well as other comparisons of classical and
squashed light, are postponed until Sec. V. Again note that for
classical noise the evolution of the three-level atom has the
same form as (\ref{squeeze}), except $M$ is restricted by $0\leq |M|\leq N$.

\subsection{Twin-beam quadrature correlations}

From the correlations between the field operators (\ref{corr}),
we can find the corresponding correlations between the quadratures.
The $X$ and $Y$ quadratures are defined as in Eq.~(\ref{XYdef}).
Then if we define the linear combinations
\beq
X^{\pm}=\frac{1}{\sqrt{2}}(X_{2}\pm X_{1}),~~Y^{\pm}=\frac{1}{\sqrt{2}}(Y_{2}\pm Y_{1}),
\eeq
we obtain the following expressions for their non-zero correlations
\bqa
\langle\tilde{X}^{\pm}(\omega)X^{\pm}(0)\rangle&=&1+2N\pm 2M, \\
\langle\tilde{Y}^{\mp}(\omega)X^{\mp}(0)\rangle&=&1+2N\pm 2M,
\eqa
where we have again set $N_{1}=N_{2}=N$ and $M$ real.

These correlations therefore correspond to the $X$ and $Y$ spectra
of the operator combinations $\{b_{2}(t)\pm b_{1}(t)\}/\sqrt{2}$ [see
Eq.~(\ref{specdef})]. Thus for twin-beam squeezing the squeezed and
stretched (depending on the sign of $M$) spectra are
\bqa
S_{X^{+}}(\omega)&=&S_{Y^{-}}(\omega)=1+2N+2M, \label{sqspec} \\
S_{Y^{+}}(\omega)&=&S_{X^{-}}(\omega)=1+2N-2M.
\eqa
That is, assuming $M<0$, the combined field $(b_{2}+ b_{1})/\sqrt{2}$
has a squeezed $X$ quadrature, while $(b_{2}- b_{1})/\sqrt{2}$ has a
squeezed $Y$ quadrature.

\section{Twin-beam squashing}

This section develops the critical result of this paper: the master
equation of a cascade three-level atom interacting with a squashed
twin-beam. This puts us in a position to compare this evolution
to that of twin-beam squeezed light and a twin-beam with classical
correlations.

\subsection{Generation of twin-beam squashed light}

Consider the double feedback schematic shown in Fig.~3, but for
now ignore the three-level atom. The light beams are shown as
double arrows because they represent twin-beams, which are again
defined by their continuum annihilation operators, i.e.
$(b_{1},b_{2})$. As for twin-beam squeezing the two fields of the
squashed twin-beam are noise correlated, where this correlation is
now produced by the feedback mechanism (detailed in this section).

Unlike squeezing, here the two fields differ by their
polarization rather than their center frequencies (now the same).
One beam $b_{1}$ has LHC polarization and the other $b_{2}$ RHC
polarization. Thus, linear combinations of the two fields can be
easily measured using homodyne detection. In principle the
combinations could still be measured if the fields have different
frequencies. The drawback in that case is the need for extremely
fast detectors whose bandwidth must exceed the frequency
difference. For the twin-beam used experimentally \cite{3laexpt95}
this would mean a detector bandwidth of the order of $10^{14}$Hz!

\begin{figure}
\includegraphics[width=0.46\textwidth]{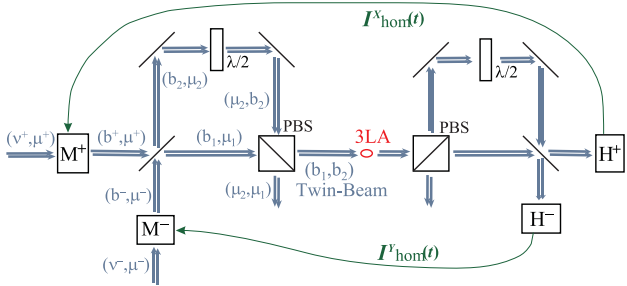}
\vspace{0.25cm} \caption{\narrowtext Experimental configuration
for a squashed twin-beam interacting with a three-level atom. The
atom is depicted in the center of the figure, coupled to the
twin-beam $(b_{1},b_{2})$ as in Fig.~2. The ``black box''
modulation and detection systems M$^{+}$, M$^{-}$, H$^{+}$,
H$^{-}$, are similar to that detailed in Fig.~1. The
$X/Y$ quadrature measured at H$^{+}$/H$^{-}$ is used to control
the modulation at M$^{+}$/M$^{-}$ and in this way produce the
squashed correlations between $b_{1}$ and $b_{2}$.}
\end{figure}

Following Fig.~3, we start on the left side with the input
fields ($\nu^{+}$,$\mu^{+}$) and ($\nu^{-}$,$\mu^{-}$). These are
both twin-beam vacuum fields, where the operators $\nu^{\pm}(t)$
and $\mu^{\pm}(t)$ each represent vacuum fluctuations as defined
by Eq.~(\ref{vacfluc}). At this point the two fields of each
twin-beam are not correlated; they are two independent vacuum
fields. Thus the quadrature operators
$X_{\nu}(t)=\nu(t)+\nu\dg(t)$ and $Y_{\nu}(t)=i\nu\dg(t)-i\nu(t)$
have statistics described by the spectra
$S_{X}(\omega)=S_{Y}(\omega)=1$.

The modulators M$^{+}$, M$^{-}$ simply add coherent amplitudes,
$\varepsilon^{+}$ and $\varepsilon^{-}$, to the LHC components of
the input vacuum fields. Both of these ``black boxes'' are the
same as the modulation region depicted in Fig.~1, or see Ref.~\cite{Wis98}
for details. The LHC fields after the modulators are
\bqa
b^{+}(t)&=&\nu^{+}(t)+\varepsilon^{+}(t), \label{b+} \\
b^{-}(t)&=&\nu^{-}(t)+i\varepsilon^{-}(t), \label{b-}
\eqa
where $\varepsilon^{+}(t)$ and $\varepsilon^{-}(t)$ are real
functions of time.

The twin-beams now enter a series of optics (a 50:50 beam splitter,
perfect mirrors, a 1/2 wave plate and a polarizing beam splitter)
to give the desired twin-beam $(b_{1},b_{2})$. The relationships
between the continuum annihilation operators $b^{\pm}$ and $b_{i}$
are then
\beq
b_{1}=\frac{1}{\sqrt{2}}(b^{+}-b^{-}),~ b_{2}=\frac{1}{\sqrt{2}}(b^{+}+b^{-}).
\label{abrel}
\eeq
It is the purpose of the feedback system to produce non-classical
correlations between $b_{1}$ and $b_{2}$, i.e. to produce the
\textit{squashed} twin-beam $(b_{1},b_{2})$ that will be coupled
to the three-level atom.

In the previous section, III.B, on \textit{squeezed} twin-beams,
equation (\ref{sqspec}) shows that the $X$ quadrature of
$(b_{1}+b_{2})/\sqrt{2}$ and the $Y$ quadrature of
$(b_{2}-b_{1})/\sqrt{2}$ are squeezed for $M<0$, i.e.
$S_{X^{+}}=S_{Y^{-}}<1$. For the generation of our
\textit{squashed} twin-beam we require the same quadratures to
have \textit{squashed} correlations. From the above relationship
(\ref{abrel}), it can be seen that the $X$ quadrature of $b^{+}$
and the $Y$ quadrature of $b^{-}$ give these combinations.

Therefore, we reverse the optical setup on the other side of the
atom. In the absence of the atom, the same fields $b^{+}(t),$
$b^{-}(t)$ will enter the homodyne detection devices H$^{+}$,
H$^{-}$, which are set up to measure the $X$ quadrature of $b^{+}$
and the $Y$ quadrature of $b^{-}$. These ``black box'' devices are
again the same as that depicted in Fig.~1. Note that M$^{+}$
and H$^{+}$ (and M$^{-}$ and H$^{-}$) are really coupled by a
shared local oscillator, which in this case is a large LHC polarized
coherent field.

Again assuming perfect detection, we obtain for the homodyne
photocurrents
\bqa
I_{\rm hom}^{X}(t) &=& X_{\nu}^{+}(t)+2\varepsilon^{+}(t), \label{IhomX} \\
I_{\rm hom}^{Y}(t) &=& Y_{\nu}^{-}(t)+2\varepsilon^{-}(t). \label{IhomY}
\eqa
Through the two feedback loops, these currents may be used to
determine the classical field amplitudes $\varepsilon^{+}(t)$
and $\varepsilon^{-}(t)$:
\bqa
\varepsilon^{+}(t)&=&\frac{g}{2}\int^{\infty}_{0}h(s)I_{hom}^{X}(t-\tau-s)ds, \label{eps+}\\
\varepsilon^{-}(t)&=&\frac{g}{2}\int^{\infty}_{0}h(s)I_{hom}^{Y}(t-\tau-s)ds \label{eps-},
\eqa
where we have assumed the same characteristics for both feedback
loops.

Solving in the Fourier domain and substituting into equations
(\ref{b+}) and (\ref{b-}) yields
\bqa
\tilde{b}^{+}(\omega) &=& \tilde{\nu}^{+}(\omega)
+\frac{g\tilde{h}(\omega)e^{-i\omega\tau}\tilde{X}_{\nu^{+}}(\omega)}{2[1-g\tilde{h}(\omega)e^{-i\omega\tau}]},
\label{b+feed}
\\
\tilde{b}^{-}(\omega) &=& \tilde{\nu}^{-}(\omega)
+i\frac{g\tilde{h}(\omega)e^{-i\omega\tau}\tilde{Y}_{\nu^{-}}(\omega)}{2[1-g\tilde{h}(\omega)e^{-i\omega\tau}]}.
\label{b-feed}
\eqa
The $X$ quadrature spectrum of $b^{+}$ and the $Y$ quadrature
spectrum of $b^{-}$ are then
\bqa
S_{X^{+}}(\omega)&=&|1-g\tilde{h}(\omega)e^{-i\omega\tau}|^{-2}, \\
S_{Y^{-}}(\omega)&=&|1-g\tilde{h}(\omega)e^{-i\omega\tau}|^{-2}.
\eqa
Assuming again the limit of broadband feedback, we set
$\tilde{h}(\bar{\omega})e^{-i\bar{\omega}\tau}=1$ to give
\beq
S_{X^{+}}(\bar{\omega})=S_{Y^{-}}(\bar{\omega})=(1-g)^{-2}.
\label{specX+Y-}
\eeq
As before these spectra are below the standard quantum limit for
negative feedback, i.e. for $g<0$.

Therefore, the $X$ quadrature of $(b_{1}+b_{2})/\sqrt{2}$ and the
$Y$ quadrature of $(b_{2}-b_{1})/\sqrt{2}$ are \textit{squashed}
as required, and we can now model the effects produced by a squashed
twin-beam $(b_{1},b_{2})$ interacting with a three-level atom.

\subsection{Inclusion of the three-level atom}

Return to Fig.~3 and now include the three-level atom. Assume for
simplicity that it is possible to mode-match all of
$(b_{1},b_{2})$ into the atom's input. Also, assume that the three
energy levels, $\ket{1}$, $\ket{2}$ and $\ket{3}$, are equally
spaced and have magnetic quantum numbers, $m_{j}$, of 0, 1, and 0
respectively. This means that $b_{1}$ (with LHC polarization) will
couple to the $\ket{1}\rightarrow\ket{2}$ transition and $b_{2}$
(with RHC polarization) will couple to the
$\ket{2}\rightarrow\ket{3}$ transition, as shown in Fig.~2.

The interaction Hamiltonian is the same as that given for
twin-beam squeezing (\ref{intH}). From here on we will set
$\gamma_{1}=\gamma_{2}=1$. Under a linear coupling as in
(\ref{intH}), $b_{1}'(t)=b_{1}(t)+s_{1}(t)$ and
$b_{2}'(t)=b_{2}(t)+s_{2}(t)$. Thus the the homodyne
photocurrents (\ref{IhomX}) and (\ref{IhomY}) become
\bqa
I_{\rm hom}^{X}(t)+\frac{1}{\sqrt{2}}\sq{s_{1}^{x}(t)+s_{2}^{x}(t)}, \\
I_{\rm hom}^{Y}(t)+\frac{1}{\sqrt{2}}\sq{s_{1}^{y}(t)-s_{2}^{y}(t)},
\eqa
where $s^{x}=s+s\dg$ and $s^{y}=is-is\dg$.

Incorporating the feedback as before we solve (\ref{eps+}) and
(\ref{eps-}) in the Fourier domain. We obtain similar expressions
for $\tilde{b}^{+}(\omega)$ and $\tilde{b}^{-}(\omega)$ as those
given in Eqs.~(\ref{b+feed}, \ref{b-feed}). The atom-field coupling
simply adds the term
$[\tilde{s}_{1}^{x}(\omega)+\tilde{s}_{2}^{x}(\omega)]/\sqrt{2}$ to
$\tilde{X}_{\nu^{+}}(\omega)$ and the term
$[\tilde{s}_{1}^{y}(\omega)-\tilde{s}_{2}^{y}(\omega)]/\sqrt{2}$ to
$\tilde{Y}_{\nu^{-}}(\omega)$. Note that the terms
$\tilde{\nu}^{+}(\omega)$ and $\tilde{\nu}^{-}(\omega)$
are unaffected by this substitution.

In the limit of broadband feedback (refer to Section II.B. for
the details of the Markov approximation), the field operators
acting on the atom are then
\bqa
b_{1}(t)&=& \sq{\nu^{+}(t)-\nu^{-}(t)}/\sqrt{2}
\nl{+} \frac{g\cu{X_{\nu^{+}}(t^{-})+[s_{1}^{x}(t^{-})+s_{2}^{x}(t^{-})]/\sqrt{2}}}{2(1-g)\sqrt{2}} \nn \\
\nl{-} i\frac{g\cu{Y_{\nu^{-}}(t^{-})+[s_{1}^{y}(t^{-})-s_{2}^{y}(t^{-})]/\sqrt{2}}}{2(1-g)\sqrt{2}},
\label{b1}
\\
b_{2}(t)&=& \sq{\nu^{+}(t)+\nu^{-}(t)}/\sqrt{2}
\nl{+} \frac{g\cu{X_{\nu^{+}}(t^{-})+[s_{1}^{x}(t^{-})+s_{2}^{x}(t^{-})]/\sqrt{2}}}{2(1-g)\sqrt{2}} \nn \\
\nl{+} i\frac{g\cu{Y_{\nu^{-}}(t^{-})+[s_{1}^{y}(t^{-})-s_{2}^{y}(t^{-})]/\sqrt{2}}}{2(1-g)\sqrt{2}},
\label{b2}
\eqa
where the relation (\ref{abrel}) has been used. Here the time
argument $t^{-}$ is again used to indicate that the fed-back
signal is infinitesimally delayed from the vacuum noise.

The total interaction Hamiltonian can be written
\beq
H(t)=H_{0}(t)+H_{\rm fb}(t), \label{totH}
\eeq
where $H_{0}(t)$ is the same as in (\ref{intH}) assuming no
feedback, that is
\bqa
H_{0}(t)&=& \frac{i}{\sqrt{2}}\sq{\nu^{+}(t)-\nu^{-}(t)}\dg s_{1}(t)+{\rm H.c}
\nl{+} \frac{i}{\sqrt{2}}\sq{\nu^{+}(t)+\nu^{-}(t)}\dg s_{2}(t)+{\rm H.c}.
\eqa
Substituting $b_{1}(t)$ and $b_{2}(t)$ into (\ref{intH}),
we find for the feedback Hamiltonian
\beq
H_{\rm fb}(t)= F_{X}(t)X'_{\nu^{+}}(t^{-})+F_{Y}(t)Y'_{\nu^{-}}(t^{-}),
\label{fbH}
\eeq
where the quadrature operators are given by
\bqa
X'_{\nu^{+}}(t^{-})&=& X_{\nu^{+}}(t^{-})+\sq{s_{1}^{x}(t^{-})+s_{2}^{x}(t^{-})}/\sqrt{2} , \nn \\
Y'_{\nu^{-}}(t^{-})&=& Y_{\nu^{-}}(t^{-})+\sq{s_{1}^{y}(t^{-})-s_{2}^{y}(t^{-})}/\sqrt{2},
\eqa
and the feedback operators are
\bqa
F_{X}(t)&=&\frac{\lambda}{2} \sq{s_{1}^{y}(t)+s_{2}^{y}(t)}/\sqrt{2}, \nn \\
F_{Y}(t)&=&\frac{\lambda}{2} \sq{s_{2}^{x}(t)-s_{1}^{x}(t)}/\sqrt{2}.
\eqa
Here we have again used the feedback parameter $\lambda$,
defined in Eq.~(\ref{lambda}).
The infinitesimal time delay denoted by $t^{-}$ ensures
that the quadrature operators $X'_{\nu^{+}}(t^{-})$ and
$Y'_{\nu^{-}}(t^{-})$ commute with all system operators
at the same time $t$ \cite{Wis94}. Therefore they will commute with
$F_{X}(t)$ and $F_{Y}(t)$ and there is no ordering
ambiguity in (\ref{fbH}).

The evolution of an arbitrary system operator $r(t)$
generated by (\ref{fbH}) is
\bqa
\sq{dr}_{\rm fb}&=&U\dg_{\rm fb}rU_{\rm fb}-r \nn \\
&=& iX'_{\nu^{+}}(t^{-})dt\sq{F_{X},r}-\frac{1}{2}\sq{F_{X},\sq{F_{X},r}}dt
\nl{+} iY'_{\nu^{-}}(t^{-})dt\sq{F_{Y},r}-\frac{1}{2}\sq{F_{Y},\sq{F_{Y},r}}dt.
\eqa
where all time arguments are $t$ unless otherwise given,
and the unitary evolution operator is defined as
$U_{\rm fb}={\rm exp}[-iH_{\rm fb}dt]$.

Adding in the non-feedback evolution gives the total Langevin
equation $dr$, which following the process of \cite{Wis94} leads
to total master equation for the atom. This involves separating
$X'_{\nu^{+}}(t^{-})dt$ such that $\nu'^{+}(t^{-})dt$ (which
equals $\nu^{+}(t^{-})dt+[s_{1}(t^{-})+s_{2}(t^{-})]dt/\sqrt{2}$ )
is moved to the right side and $[\nu'^{+}(t^{-})]\dg dt$ is moved
to the left side of $[F_{X},r]$ (using the fact that $\nu'^{\pm}(t^{-})$
commute with system operators at the same time). The same is done
for $Y'_{\nu^{+}}(t^{-})$.

The operators $\nu'(t^{-})$ are now in the correct position to
annihilate the vacuum when the trace over the bath density
operator is taken. Taking the total trace over the system and bath
density operators gives $\an{dr}$, and here we can ignore the time
difference since for arbitrary system operators $a(t)$ and $b(t)$
\beq
{\rm lim}_{\tau\rightarrow 0}\an{a(t)b(t-\tau)}=\an{a(t)b(t)}.
\eeq
The final step involves changing from the \hei to the \sch
picture and deriving the evolution of the density operator
for the atom alone. As noted previously, this is found from
Eq.~(\ref{HP-SP}).

The resultant master equation for a cascade three-level atom
interacting with twin-beam squashed light is
\bqa
\dot{\rho} &=& \ro{1+\lambda+\frac{\lambda^{2}}{4}}\cu{ {\cal D}[s_{1}]\rho + {\cal D}[s_{2}]\rho }
\nl{+} \frac{\lambda^{2}}{4} \{ {\cal D}[s_{1}\dg]\rho + {\cal D}[s_{2}\dg]\rho \}
\nl{+} \frac{\lambda}{2} \cu{\sq{s_{1},\sq{s_{2},\rho}}+[s\dg_{1},[s\dg_{2},\rho]]}
\nl{+} \frac{\lambda^{2}}{8} \left\{ \sq{s_{1},\sq{s_{2},\rho}}+[s\dg_{1},[s\dg_{2},\rho]] \right. \nn \\
&&~~~~~~~~~~ \left. +~\sq{s_{2},\sq{s_{1},\rho}}+[s\dg_{2},[s\dg_{1},\rho]] \right\}.
\label{squash}
\eqa

The spectra for the squashed quadratures were given in
Eq.~(\ref{specX+Y-}). We can see that in terms of the feedback
parameter $\lambda$ they have the same form as the squashed
spectra for the single squashed beam in Section II.C. This is
given by Eq.~(\ref{speclamb}). Note that the spectra are only
squashed for $-1<\lambda<0$. These quadrature correlations can be
re-expressed in terms of the field operators $b_{1}$ and $b_{2}$.
We find that the non-zero correlations for squashing are ($i\neq j$)
\bqa
\langle b_{i}\dg(t) b_{i}(t') \rangle &=&\frac{\lambda^{2}}{4}\delta(t-t'), \nn \\
\langle b_{i}(t)b\dg_{i}(t') \rangle &=&\ro{1+\lambda+\frac{\lambda^{2}}{4}}\delta(t-t'), \nn \\
\langle b_{i}(t)b_{j}(t') \rangle &=&\ro{\frac{\lambda}{2}+\frac{\lambda^{2}}{4}}\delta(t-t'), \nn \\
\langle b\dg_{i}(t)b\dg_{j}(t') \rangle &=&\ro{\frac{\lambda}{2}+\frac{\lambda^{2}}{4}}\delta(t-t').
\label{corrlamb}
\eqa
Again the $\delta$-functions here are not exact. Their width is of the
order $B^{-1}$ where $B\gg 1$ is the squashing bandwidth. Note the modified
commutation relations implied by the first two equations here, i.e.
$[b_{i}(t),b_{i}\dg(t')]\neq\delta(t-t')$.

If we had constructed our feedback mechanism such that we squashed
the $Y^{+}$ and $X^{-}$ quadratures rather than $X^{+}$ and
$Y^{-}$, then the correlations would be same as in
(\ref{corrlamb}) except $\langle b_{i}(t)b_{j}(t') \rangle$ and
$\langle b\dg_{i}(t)b\dg_{j}(t') \rangle$ would be negative.
Correspondingly, the master equation for the atom would be the
same as Eq.~(\ref{squash}) except the last two terms ($\lambda/2$
and $\lambda^{2}/8$) would be negative.

\section{Comparison of squashed light with squeezed and classical light}

This section collects the main results of the previous sections,
which are the correlations of squeezed, squashed and classical
light, and the resulting master equations due to the interaction
of this light with two- and three-level atoms. We find that that a
unified formalism can be found in terms of effective $N$ and $M$
parameters. The master equation for squashed light is then
compared to those for squeezed and classical light in order to
determine the non-classical nature of squashed light.

\subsection{Unified formalism}

The general formalism for broadband squeezed, squashed and classical
light interacting with simple atomic systems involves effective $N$
and $M$ parameters. For squeezed and classical fields, these values
are the usual broadband limits of the photon number function and the
two-photon correlation function. For squashed fields it is more
complicated. Due to the fact that it is not a free field (see Section II.B.),
the correlations of squashed light need to be separated into upward and
downward processes.

The most general parameters for the three types of light are
$N_{U}$, $N_{D}$, $M_{U}$ and $M_{D}$, where the $U$ and $D$ subscripts
refer to upward (excitation) and downward (de-excitation) processes.
The two-level atom master equations for squeezing (\ref{2LAsqueeze})
and squashing (\ref{2LAsquash}) can both be rewritten in terms of these
parameters to give the same general master equation. It is
\bqa
\dot{\rho} &=& (1+N_{D}){\cal D}[\sigma]\rho + N_{U}{\cal D}[\sigma\dg]\rho
\nl{-} \frac{1}{2}\ro{M_{D}+M_{U}}\cu{ \sigma\rho\sigma +\sigma\dg\rho\sigma\dg }.
\label{2LA}
\eqa
For the three-level atom, the general master equation is obtained
by unifying Eqs.~(\ref{squeeze}) and (\ref{squash}). The result is
\bqa
\dot{\rho}&=& (1+N_{D})\{ {\cal D}[s_{1}]\rho + {\cal D}[s_{2}]\rho \}
+ N_{U}\{ {\cal D}[s_{1}\dg]\rho + {\cal D}[s_{2}\dg]\rho \}
\nl{+} \frac{1}{2}M_{D} \cu{\sq{s_{1},\sq{s_{2},\rho}}+[s\dg_{1},[s\dg_{2},\rho]]}
\nl{+} \frac{1}{2}M_{U} \cu{\sq{s_{2},\sq{s_{1},\rho}}+[s\dg_{2},[s\dg_{1},\rho]]}.
\label{3LA}
\eqa

The effective $N$ and $M$ parameters are the same for both
master equations. For squashed light they are
\bqa
N_{U}&=&\lambda^{2}/4,~~~N_{D}=\lambda+\lambda^{2}/4, \nn \\
M_{U}&=&\pm N_{U},~~M_{D}=\pm N_{D},
\label{sqshparam}
\eqa
where $-1<\lambda<0$. Here the $+$ refers to measuring the
$X^{+}$ and $Y^{-}$ quadratures and vice versa for the $-$,
as explained at the end of the preceding section.
For maximally squeezed light we have
\bqa
N_{U}&=& N_{D} = N, \nn \\
M_{U}&=& M_{D} = \mp\sqrt{N(N+1)},
\label{sqzeparam}
\eqa
and for maximally correlated classical light
\bqa
N_{U}&=& N_{D} = N, \nn \\
M_{U}&=& M_{D} = \mp N.
\label{clasparam}
\eqa

In all three cases, the sign of $M$ indicates which quadratures are
squashed (below-shot-noise spectra), squeezed (below-shot-noise spectra)
or classically noisy (above-shot-noise spectra).
For the single field the $X$ and $Y$ quadrature spectra are
\bqa
S_{X}(\omega)&=&1+N_{U}+N_{D}+M_{U}+M_{D}, \label{Sx}\\
S_{Y}(\omega)&=&1+N_{U}+N_{D}-M_{U}-M_{D},
\eqa
while for the twin-beam we have
\bqa
S_{X^{+}}(\omega)&=&S_{Y^{-}}(\omega)=1+N_{U}+N_{D}+M_{U}+M_{D},\label{Sx+Sy-} \\
S_{X^{-}}(\omega)&=&S_{Y^{+}}(\omega)=1+N_{U}+N_{D}-M_{U}-M_{D}.
\eqa

Thus, to clarify the above equations, if $M_{U/D}=+N_{U/D}$ for
squashed light, then $S_{X}=S_{X^{+}}=S_{Y^{-}}<1$ (since
$\lambda<0$) while the conjugate quadratures
$S_{Y}=S_{Y^{+}}=S_{X^{-}}=1$. If $M_{U/D}=-N_{U/D}$ the reverse
is true. These results are the same for classical light except now
the affected spectra are above unity, i.e. if $M=N$ then
$S_{X}=S_{X^{+}}=S_{Y^{-}}>1$ while $S_{Y}=S_{Y^{+}}=S_{X^{-}}=1$,
and vice versa for $M=-N$. For squeezed light we have the expected
Heisenberg trade-off. That is, if $M=\sqrt{N(N+1)}$ then
$S_{X}=S_{X^{+}}=S_{Y^{-}}>1$ while $S_{Y}=S_{Y^{+}}=S_{X^{-}}<1$,
and vice versa for $M=-\sqrt{N(N+1)}$.

\subsection{Effective \textit{N} and \textit{M} parameters}

The values for $N_{U}$, $N_{D}$ and $M_{\rm av}=(M_{U}+M_{D})/2$
can be simply related to the field correlations. For a single
broadband squashed (or squeezed or classical) field these are
\bqa
&\langle b\dg(t)b(t')\rangle&=N_{U}\delta(t-t'),\nn \\
&\langle b(t)b\dg(t')\rangle&=(1+N_{D})\delta(t-t'),\nn \\
&\langle b(t)b(t')\rangle&=\langle b\dg(t)b\dg(t')\rangle=M_{\rm av}\delta(t-t'),
\eqa
The unified correlations for the twin-beam combine
Eqs.~(\ref{corr}) and (\ref{corrlamb}) and are given by ($i\neq j$)
\bqa
\langle b_{i}(t)\dg b_{i}(t') \rangle &=&N_{U}\delta(t-t'), \nn \\
\langle b_{i}(t)b\dg_{i}(t') \rangle &=&(1+N_{D})\delta(t-t'), \nn \\
\langle b_{i}(t)b_{j}(t')\rangle  &=&\langle b\dg_{i}(t)b\dg_{j}(t') \rangle=M_{\rm av}\delta(t-t').
\eqa

The individual choices for $M_{U}$ and $M_{D}$ are made apparent
when the master equations for the three-level atom is broken into
upwards and downwards transitions. This is best illustrated by
expressing the general master equation in diagrammatic form.
Fig.~4 shows the diagrammatic master equation for the three-level
atom, where $\rho_{13}$ (which equals $\rho_{31}$ for real $M$) is
the two-photon coherence between the upper and lower states.

\begin{figure}
\hspace{1cm}
\includegraphics[width=0.3\textwidth]{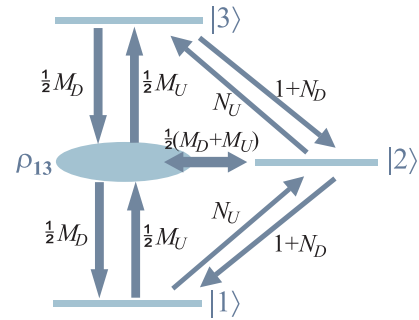}
\vspace{0.25cm} \caption{\narrowtext Diagrammatic master equation
for a cascade three-level atom interacting with a broadband field.
The levels are labelled by $\ket{1}$, $\ket{2}$ and $\ket{3}$,
while the two-photon coherence is given by $\rho_{13}$. The
transition rates, indicated by the arrows, are given by the
effective $N$ and $M$ parameters.}
\end{figure}

The transition rates in this diagram are found by re-expressing
the master equation in terms of the state vectors $\ket{1}$,
$\ket{2}$ and $\ket{3}$ and using the matrix representation for
the density operator. That is
\beq
\rho=\sum^{3}_{i,j=1}\rho_{ij}\ket{i}\bra{j},
\eeq
where the matrix elements for $i=j$ are the occupation
probabilities and those for $i\neq j$ are the coherences. We find
two closed sets of coupled differential equations for $\rho_{ij}$.
The first shows that $\rho_{12}=\rho_{21}=\rho_{23}=\rho_{32}=0$.
The second set contains the equations of motion for the three
populations $\rho_{ii}$ and the two-photon coherence $\rho_{13}$.
These coupled equations give the transition rates seen in Fig.~4.

By simply comparing the correlation parameters, squashed light
appears classical in the sense that $|M|=N$. Yet the squashed
spectra, e.g. $S_{X}=1+2\lambda+\lambda^{2}$, are below the
standard quantum limit of unity for $-1<\lambda<0$ which is
clearly non-classical. For the spectra, terms involving $\lambda$
can thus be thought of as the non-classical terms, while terms
involving $\lambda^{2}$ are classical.

Looking at the master equation (see Fig.~4) we see that this
non-classical term also appears in the de-excitation process since
$N_{D}=M_{D}=\lambda+\lambda^{2}/4$, whereas the excitation
process appears classical in all regards since here
$N_{U}=M_{U}=\lambda^{2}/4$. Therefore, we can anticipate that the
non-classical two-photon excitation seen for squeezed light might
not be seen for squashed light. However, we can also expect
there to be a corresponding non-classical process in the
de-excitation.

\subsection{Basis for comparison}

The features of the master equations for squashed, squeezed and
classical light can now be compared. Here we address the question
of what to keep constant between these three cases. In the past,
squeezed and classical light have been compared in terms of the
intensity since it is defined in the same way for both types of
light. The most obvious definition for the intensity ${\cal N}$ is
in terms of the photon flux operator, i.e.
\beq
\langle b\dg_{i}(t)b_{i}(t') \rangle ={\cal N}\delta(t-t').
\label{N}
\eeq
If we extend this definition of
intensity to include squashed light, we must not forget that, in
this case, commuting the operators does not change ${\cal N}$ into
$1+{\cal N}$. Therefore the term ``intensity'' does not have the
same meaning for squashed light. Nevertheless, the parameter
${\cal N}$ is the most natural choice as we will see below.

Applying this definition of intensity (\ref{N}) for all
three types of light we see that, in general, ${\cal N}=N_{U}$.
For the squeezed and classical fields this simply means that
$N={\cal N}$ and $|M|=\sqrt{{\cal N}({\cal N}+1)}$ for squeezing,
and $|M|={\cal N}$ for classical light. For squashed light the
relationship is $\lambda=-2\sqrt{\cal N}$. Since the spectra
(\ref{Sx}) and (\ref{Sx+Sy-}) are only below unity for
$-1<\lambda<0$ (negative feedback), the ``intensity'' range for
squashed light is thus limited to $0<{\cal N}<0.25$.

It is interesting to note that the feedback described in Sections
II.B and IV can be modified to produce classical light which has
the same intensity as the squashed light. This would enable
experimental comparison of squashed and classical light to be made
with the same apparatus. As opposed to using positive feedback
$g>0$, which increases the noise in the fed-back quadratures, the
method introduced here (see Fig.~5) actually produces maximally
correlated classical light, i.e. $N_{U}=N_{D}=N$ and
$M_{U}=M_{D}=N$.

Consider the simplified diagram for producing squashed light seen
in Fig.~5(a). This theory is detailed in \cite{Wis98,Wis99} and is
also outlined in Section II.B. Now, by simply separating the
feedback loop from the test system, as shown in Fig.~5(b), it can
be shown that the light, $a(t)$, interacting with the system is then
maximally correlated classical noise. Note that it is a relatively
simple matter to extend this for the twin-beam case.

\begin{figure}
\hspace{1cm}
\includegraphics[width=0.3\textwidth]{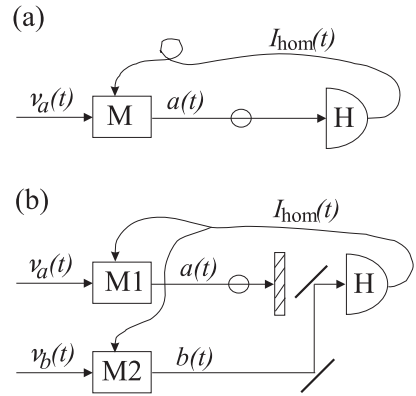}
\vspace{0.25cm} \caption{\narrowtext Simplified diagrams for (a)
the production of squashed light and (b) the production of
maximally correlated classical light using feedback. The circles
are where a test atom could be positioned.}
\end{figure}

Here we start with two initial vacuum fields $\nu_{a}$ and
$\nu_{b}$ to which coherent amplitudes are again added (by M1
and M2) resulting in the fields $a$ and $b$. However, only
the $X$ quadrature of $b$ is measured and used to control
the same coherent amplitude, $\varepsilon$, added to both the
initial vacuum fields. That is, $a=\nu_{a}+\varepsilon$ and
$b=\nu_{b}+\varepsilon$. Assuming the same modulation as in
Eq.~(\ref{epsilon}), we obtain in the broadband approximation
\beq
a(t)=\nu_{a}(t)+\frac{gX_{b}(t^{-})}{2(1-g)}. \label{a}
\eeq

Coupling a two-level atom this field with perfect mode-matching
again leads to the general master equation (\ref{2LA})
except now
\bqa
N_{U}&=& {\cal N} =N_{D}= \lambda^{2}/4 \nn \\
M_{U}&=& M_{D} = \lambda^{2}/4.
\eqa
Therefore, by comparison with Eqs.~(\ref{clasparam}), we see that
this is maximally correlated classical light with the same
intensity as squashed light (\ref{sqshparam}).

Note that here $M$ is positive which means that $S_{X}>1$ and
$S_{Y}=1$, which can also be noted from the fact that we are
adding noise to the $X$ quadrature of $a$ (\ref{a}). To set
$S_{X}=1$ and $S_{Y}>1$ we would need to measure the $Y$
quadrature of $b$ and use it to add noise to the $Y$ quadrature of $a$.

By choosing a fixed value for ${\cal N}$ we are able to make a
phase space comparison between squashed, squeezed and classical
light. This is shown in Fig.~6 for ${\cal N}=0.1$, where we have
chosen $M_{U/D}=N_{U/D}$ for squashed light and $M<0$ for squeezed
and classical light. We emphasize that these are not the contours
of a Wigner function of a single-mode harmonic oscillator. All the
fields discussed in this paper are multi-mode fields denoted by
continuum annihilation operators (see Secs. II.A and III.A). In
particular, squashing has no meaning for a single-mode field.
Figure~6 is best interpreted as a radial plot, where the distance
of the curve from the origin at an angle $\theta$ represent the
broadband values of the quadrature spectra,
$S_{Q_{\theta}}(\bar{\omega})$.

\begin{figure}
\hspace{0.7cm}
\includegraphics[width=0.35\textwidth]{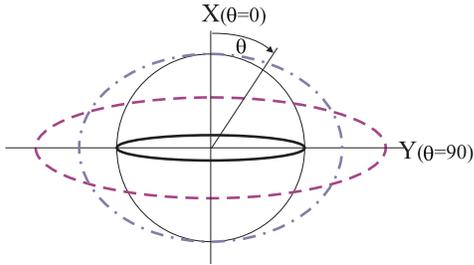}
\vspace{0.25cm} \caption{\narrowtext Phase space comparison
between squashed, squeezed and classical light, where the
intensity ${\cal N}=0.1$ is kept fixed for all three cases. The
contours represent the value of $S_{Q_{\theta}}(\bar{\omega})$, where
$Q_{\theta}=X\cos\theta+Y\sin\theta$. The circle represents the vacuum
state, the solid ellipse is the squashed state, the dashed ellipse
is the squeezed state and the dash-dot ellipse represents the
classical noise state.}
\end{figure}

This figure highlights the difference between squashed and
squeezed light that was discussed previously in Sec. II.B. While
both are non-classical fields in the sense of reduced fluctuations
in the $X$ quadrature, squashed light does not have the
corresponding increase in the $Y$ quadrature. Hence, the
fluctuations are simply \textit{squashed} from the vacuum limit in
one quadrature, as opposed to \textit{squeezed} from one
quadrature into the other.

For the rest of the paper we will assume correlations like that
given in Fig.~6. That is, $M_{U/D}=N_{U/D}$
($S_{X}=S_{X^{+}}=S_{Y^{-}}<1$) for squashed light,
$M=-\sqrt{N(N+1)}$ ($S_{X}=S_{X^{+}}=S_{Y^{-}}<1$) for squeezed
light, and $M=-N$ ($S_{X}=S_{X^{+}}=S_{Y^{-}}=1$) for classical
light.

\subsection{Steady state excited populations}

The squeezed master equation leads to a steady state excited
population with a linear (non-classical), as well as quadratic
(classical), dependence on the intensity. This has been shown
both theoretically \cite{FicDru91,FicDru97} and experimentally
\cite{3laexpt95}.

Normal (classical) two-photon excitation involves a two-step
process with one photon absorbed in each step (hence the quadratic
dependence). Squeezed light has large two-photon correlations and
therefore the excitation from the ground to the upper state can
occur in one step. We have shown that squashed light also has
non-classical two-photon correlations as evidenced by the
below-shot-noise spectrum (\ref{speclamb}). However, squashed
light does not produce an excited population with a linear
dependence on the intensity, as shown below.

The steady state population of the upper state for the twin-beam
squeezed vacuum field is
\beq
\rho_{33}=\frac{\cal N}{1+2{\cal N}}\simeq{\cal N}.
\label{p33Ns}
\eeq
The corresponding result for a twin-beam with classical noise is
\beq
\rho_{33}=\frac{2{\cal N}^{2}}{(1+2{\cal N})(1+3{\cal N})}\simeq 2{\cal N}^{2}.
\label{p33Nc}
\eeq
Here the asymptotic values are for the low intensity regime,
${\cal N}\ll 1$.
The first equation shows the linear dependence on ${\cal N}$ that
is characteristic of squeezed light, whereas the corresponding
excited population for classical light has a quadratic dependence
on ${\cal N}$. The exact equation for squeezed light
also has a quadratic dependence on ${\cal N}$ for imperfect
mode-matching (not included in our analysis).

The excited population for twin-beam squashed light is given by
\beq
\rho_{33}=\frac{2{\cal N}^{2}}{1-\sqrt{\cal N}(6+10{\cal N})+{\cal N}(13+6{\cal N})}
\simeq 2{\cal N}^{2},
\label{p33} \\
\eeq
For low intensities, $\sqrt{\cal N}\ll 1$ (which really corresponds
to weak feedback $|\lambda|\ll 1$), this equation is clearly
classical with a quadratic dependence on ${\cal N}$. Thus, we see
that squashed light fails to give a non-classical excited population
similar to squeezed light.

\subsection{Transient two-photon coherences}

We have shown that squashed light excitation of a three-level atom
is a classical process. This can be understood by looking at the
two-photon transition rate for excitation, $M_{U}=\lambda^{2}/4$.
As we pointed out earlier $\lambda^{2}$ corresponds to the classical
term in the squashed spectrum (equal to $1+2\lambda+\lambda^{2}$).
It is the two-photon \textit{de}-excitation rate that has the
non-classical dependence on $\lambda$, i.e. $M_{D}=\lambda+\lambda^{2}/4$.

This section will show that squashed light de-excitation of a
three-level atom is non-classical. However, it only behaves
similarly to squeezed light early during the de-excitation and
then only for weak feedback. This can be seen in the transient
coherence $\rho_{13}$ between the upper and lower levels.
In the short time regime, $t\ll 1$, the coherence will have only
received population from the initial state, $\ket{3}$.
For all three types of light (squashed, squeezed and classical),
the transient coherence from the upper level is
\beq
\rho_{13}\simeq\frac{M_{D}}{2}t
\eeq
which can be seen in Fig.~4.

To compare the coherences in terms of the intensity of the input
light (squashed, squeezed or classical) we simply substitute the
expressions for $M_{D}$ in terms of the respective intensities.
For squashed light this means setting $M_{D}=-2\sqrt{\cal N}+{\cal N}$,
while for maximally squeezed light $M_{D}=-\sqrt{{\cal N}({\cal N}+1)}$,
and for maximally correlated classical light $M_{D}=-{\cal N}$.
The resultant transient coherences are
\bqa
{\rm Squashed:}&~~&\rho_{13}\simeq-\sqrt{\cal N}t, \label{p13dn} \\
{\rm Squeezed:}&~~&\rho_{13}\simeq-\frac{\sqrt{\cal N}}{2}t, \label{p13sq} \\
{\rm Classical:}&~~&\rho_{13}\simeq-\frac{\cal N}{2}t. \label{p13cl}
\eqa
Again we have assumed that the intensity (or feedback strength)
is small, i.e. $\sqrt{\cal N}\ll 1$. These results clearly show that
squashed light \textit{de}-excitation is a \textit{non}-classical
process. The coherence for squashed light de-excitation (\ref{p13dn})
scales in the same non-classical way as squeezed light (\ref{p13sq}),
i.e. $\propto\sqrt{\cal N}$ \cite{posfb}.

For comparison, we can also calculate the two-photon transient
coherences for excitation. From Fig.~4 these are given by
$\rho_{13}\simeq M_{U}t/2$. For squeezed
and classical light the excitation coherences are exactly the same
as the de-excitation coherences in Eqs.~(\ref{p13sq}) and (\ref{p13cl}),
as one would expect. However, for squashed light $M_{U}\neq M_{D}$
and we obtain
\beq
{\rm Squashed~excitation:}~~\rho_{13}\simeq\frac{\cal N}{2}t,
\label{p13up}
\eeq
where again this is the weak feedback limit. We see that the
two-photon excitation for squashed light (\ref{p13up}) scales
in the same way as classical light (\ref{p13cl}), i.e. $\propto{\cal N}$.
The transient coherences thus confirm the classical nature
of squashed light excitation as indicated by the excited
populations in the preceding section.

Thus, in terms of the intensity ${\cal N}$, we see a clear scaling
difference between the classical (squashed excitation and classical
light) and non-classical (squashed de-excitation and squeezed light)
processes. For both the steady-state excited populations and the
transient two-photon coherences, the non-classical scaling is the
square root of the classical scaling.

\section{Conclusion}

We have shown in this paper that a unified formalism in terms of
effective $N$ and $M$ parameters can be found for squashed,
squeezed and classical light. By simply comparing these parameters
it appears that squashed light is classical ($M=N$). Yet there is
the contradictory, non-classical fact that it produces
photocurrents with noise below the standard quantum limit.

To better understand this unusual non-classical nature of squashed
light it is important to look in detail at its interaction with
both the two-level, and the cascade three-level atoms. By doing
so, it becomes apparent that there are differences between the
upward and downward transitions; $N_{U}=M_{U}$ are quadratically
dependent on $\lambda$, the feedback parameter, while
$N_{D}=M_{D}$ also have a linear dependence. This difference is
not present in either squeezed or classical light, and is a direct
consequence of the fact that squashed light is not a free field.

Thus, squashed light does not have a direct correspondence to
squeezed light as previous work seemed to indicate
\cite{Wis98,Wis99}. In particular, the observed experimental
signature of squeezing (non-classical excited population of a
cascade three-level atom \cite{3laexpt95,FicDru91,FicDru97}) is
not reproduced. However, the de-excitation of the three-level atom
does behave non-classically.

This is most easily seen in the transient two-photon coherence
between the highest and lowest levels of an atom prepared in the
highest level. The transient coherence scales in the same
non-classical way for both squeezing and squashing. In terms of
the intensity, ${\cal N}=N_{U}$, both the squashed and squeezed coherences
are proportional to the square root of the intensity, while
classical light has a linear dependence.

To summarize our results, we find that for squashed light,
excitation processes are classical, but \textit{de}-excitation
processes are \textit{non}-classical. This non-classical
de-excitation (lowering) is a general feature of squashed light,
as evidenced by the line narrowing of the fluorescence of a
two-level atom, and the non-classical transient coherence of
de-excitation from the upper level of a cascade three-level atom.

\end{multicols}


\begin{references}
\vspace{-1.5cm}

\bibitem{WalMil94}
D.F. Walls and G.J. Milburn, \textit{Quantum Optics} (Springer,
Berlin,1994).

\bibitem{Bucetal99}
B.C. Buchler {\em et al}, \textit{Opt. Lett.} {\bf 24}, 259
(1999).

\bibitem{Wis99}
H.M. Wiseman, \textit{J. Opt. B: Quantum Semiclass. Opt.} {\bf 1},
459 (1999).

\bibitem{why}
The term ``squashed'' refers to the reduction in fluctuations in one
quadrature of the field, with no increase in fluctuations in the other.
In squeezed light, by contrast, the fluctuations are ``squeezed'' into
the other quadrature. For a more detailed discussion see
Sec. II.B. and Fig. 6.

\bibitem{Wis98}
H.M. Wiseman, \textit{Phys. Rev. Lett.} {\bf 81}, 3840 (1998).

\bibitem{Gar86}
C.W. Gardiner, \textit{Phys. Rev. Lett.} {\bf 56}, 1917 (1986).

\bibitem{Turetal98}
Q.A. Turchette, N.Ph. Georgiades, C.J. Hood, H.J. Kimble, and A.S.
Parkins, \textit{Phys. Rev. A} {\bf 58}, 4056 (1998).

\bibitem{3laexpt95}
N.Ph. Georgiades, E.S. Polzik, K.Edamatsu, H.J. Kimble and A.S.
Parkins, \textit{Phys. Rev. Lett.} {\bf 75}, 3426 (1995).

\bibitem{FicDru91}
Z. Ficek and P.D. Drummond, \textit{Phys. Rev. A} {\bf 43}, 6247
(1991); {\em ibid.} (Part II) {\bf 43}, 6258 (1991).

\bibitem{FicDru97}
Z. Ficek and P.D. Drummond, \textit{Physics Today}, p.35, September,
1997.

\bibitem{Slu85}
R.E. Slusher {\em et al}, \textit{Phys. Rev. Lett.} {\bf 55}, 2409 (1985).

\bibitem{DalFicSwa99}
B.J. Dalton, Z. Ficek and S. Swain, \textit{J. Mod. Opt.} {\bf 46},
379 (1999).

\bibitem{CavSch85}
C.M. Caves and B.L. Schumaker, \textit{Phys. Rev. A} {\bf 31}, 3068
(1985).

\bibitem{Sha87}
J.M. Shapiro {\em et al}, \textit{J. Opt. Soc. Am. B} {\bf 4}, 1604 (1987).

\bibitem{blurb}
Here the inequality is unity rather than $1/16$, as in
Ref.~\cite{Sha87}, because our quadratures, Eq.~(\ref{XYdef}), are
defined such that $S_{X}=S_{Y}=1$ for the vacuum state.

\bibitem{Gla63}
R.J. Glauber, \textit{Phys. Rev. A} {\bf 130}, 2529 (1963); {\em
ibid.} {\bf 131}, 2766 (1963).

\bibitem{Sud63}
E.C.G. Sudarshan, \textit{Phys. Rev. Lett.} {\bf 10}, 277 (1963).

\bibitem{Gar91}
C.W. Gardiner, \textit{Quantum Noise} (Springer, Berlin, 1991).

\bibitem{MacYam86}
S. Machida and Y. Yamamoto, \textit{Opt. Commun.} {\bf 57}, 290 (1986).

\bibitem{SteSubWil90}
J.J. Stefano, A.R. Subberud and I.J. Williams, \textit{Theory and
Problems of Feedback and Control Systems 2e} (McGraw-Hill, New
York, 1990).

\bibitem{YamImoMac86}
Y. Yamamoto, N. Imoto and S. Machida, \textit{Phys. Rev. A} {\bf 33},
3243 (1986).

\bibitem{WalJak95}
J.G. Walker and E. Jakeman, \textit{Proc. Soc. Photo-Opt. Instrum. Eng.}
{\bf 492}, 274 (1995).

\bibitem{Wis94}
H.M. Wiseman, \textit{Phys. Rev. A} {\bf 49}, 2133 (1994);
Errata {\em ibid.} {\bf 49} 5159 (1994) and {\em ibid.} {\bf 50},
4428 (1994).

\bibitem{posfb}
Another interesting feature is that even for positive feedback,
i.e. $\lambda>0$, the in-loop light is not squashed (in fact it
has increased noise) but its transient, two-photon coherence for
de-excitation still has non-classical scaling. In this case, the
relationship between $M_{D}$ and the intensity is now
$M_{D}=2\sqrt{\cal N}+{\cal N}$. Hence, the two-photon coherence
$\rho_{13}$ is still proportional to $\sqrt{\cal N}$; the only
difference is the sign.

\end{references}
\end{document}